\documentclass{iopjournal}

\usepackage[backend=biber,style=vancouver,maxnames=10,minnames=10]{biblatex}
\usepackage{aas_macros}
\usepackage[utf8]{inputenc}  
\usepackage{xcolor}
\usepackage{amsmath}
\usepackage{booktabs}
\newcommand{\rev}[1]{\textcolor{black}{#1}}
\newcommand{\revv}[1]{\textcolor{black}{#1}}

\begin{document}
\articletype{Article type} 
\let\WriteBookmarks\relax
\def\floatpagepagefraction{1}
\def\textpagefraction{.001}



\title {Graph Neural Networks Trained on Null Signal for Angle Reconstruction in X-ray Polarimetry}                      



%

\author{Vittorio Latorre$^{1,*}$,
Victor Rodriguez-Fernandez$^2$,
Alessandro Di Marco$^3$,
Fabio La Monaca$^3$,
Fabio Muleri$^3$,
Paolo Soffitta$^3$ 
  }

\affil{$^1$ Universtitas Mercatorum, Piazza Mattei 10, 00186 Rome, Italy}\\
\affil{$^2$ Universidad Politécnica de Madrid, Madrid, Spain}\\
\affil{$^3$ INAF-IAPS, Via Fosso Del Cavaliere 100, 00133 Rome, Italy}\\
\affil{$^*$Author to whom any correspondence should be addressed.}
\email{vittorio.latorre@unimercatorum.it}

\begin{abstract}
Scientific real detectors often produce sparse, irregular data defined on non-Euclidean domains, where conventional convolutional neural networks (CNNs) impose geometric biases that distort physical observables. In X-ray polarimetry, such distortions can mimic polarization signals at the percent level, critically limiting measurement fidelity. We \rev{consider} a framework based on Graph Neural Networks (GNNs) for reconstructing photoelectron emission angles from Gas Pixel Detectors, operating directly on their native sparse hexagonal topology. The network is trained solely on unpolarized simulated data and integrates rotational data augmentation, ensemble averaging, and modulation-aware model selection to \rev{reduce} spurious angular modulation. Despite never encountering polarized examples, the model \rev{is able to recover} \rev{the overall structure of} polarized signals.
\rev{Compared with the classical Method of Moments (MoM), an analytical approach to track angle reconstruction, the proposed GNN achieves lower per-track error but yields worse performance for polarimetry, revealing a mismatch between local reconstruction accuracy and global polarimetric performance. Compared to CNN-based approaches, the framework shows competitive performance on both unpolarized and polarized reconstructions, despite being trained in a more challenging power-law energy distribution. These results position the method as a diagnostic tool to highlight the limitations and trade-offs of learning-based approaches for X-ray-polarimetry signal reconstruction. }
\end{abstract}





\section{Introduction}

Many detectors in physics and astrophysics produce data on irregular, sparse, and non-Euclidean domains, yet most machine learning pipelines still handle these data in formats tailored for convolutional neural networks (CNNs). This mismatch causes the imposed inductive biases of CNNs, local continuity, translation invariance, and square-grid structure to \rev{introduce} systematic distortions when detector readouts, defined on hexagons or \rev{irregular grid geometries}, are projected onto square images. For high-precision tasks, these artifacts are not minor \rev{and can introduce} spurious signals that \rev{interfere} with the true physical effect being measured. 
\rev{Moreover, detector outputs are not used in isolation but are aggregated to estimate physically relevant observables. This introduces a two-level structure in the problem: a local level, corresponding to the reconstruction of individual measurements, and a global level, corresponding to the estimation of aggregate quantities. While learning models may improve reconstruction accuracy at the local level, it remains unclear whether such improvements translate into better performance at the global level, which ultimately determines the quality of the physical measurement.}
Addressing how to design Machine Learning (ML) models that respect the native geometry and sparsity of scientific data, \rev{while also ensuring that improvements at the level of individual predictions translate into reliable estimates of aggregate physical observables,} is therefore an open challenge in many application domains.

In this paper, we focus on X-ray polarimetry among the various possible applications. In X-ray polarimetry, the polarization state of incoming photons is inferred from the angular distribution of \emph{photoelectrons} (PE), 
and the distribution of the PE emission angle directly encodes the polarization of the incident photon, following a $\cos^2(\theta)$ dependence in the differential cross section~\cite{Heitler1954}. \rev{A brief overview of these concepts is provided in Appendix \ref{app:phys}}. The Gas Pixel Detectors (GPDs) used in this field record these photoelectron tracks as sparse activations on a hexagonal pixel array, forming the basis for reconstructing the photon’s polarization state. 
Because most astrophysical sources exhibit polarization degrees of only a few percent, even small reconstruction artifacts can mimic real polarization signals, making robust angle reconstruction a critical computational requirement. 
This astrophysical application thus provides an ideal test case for evaluating advanced ML methodologies, as it presents distinctive computational challenges arising from the geometric mismatch between the detector architecture and conventional ML models.

The polarization state of an X-ray source encodes information about the geometry, magnetic fields, and emission mechanisms of extreme astrophysical systems, offering insights often inaccessible through spectral or timing analyses alone. 
Although recent missions have enabled precise polarization measurements, further improvements in reconstruction methods are needed to enhance accuracy and sensitivity. A key performance metric is the \emph{modulation factor}, defined as the modulation amplitude measured for a 100\% polarized source. For unpolarized sources, this value should ideally be zero, so any residual modulation indicates systematic reconstruction artifacts that can mimic real polarization signals. \rev{Therefore, avoiding the introduction of residual modulation is a primary requirement for any reconstruction method.}

This makes detector performance critical: advances in detector technology, particularly the development of high-granularity gas detectors~\cite{Costa2001,Bell2006,Bell2007,Baldini2021}, have enabled the measurement of the degree and angle of X-ray polarization. These measurements, which now reach the statistical precision required for astrophysical modeling~\cite{Weiss2023}, are opening new frontiers in our understanding of compact and extended galactic and extragalactic X-ray sources.

\rev{The analysis pipelines in X-ray polarimetry are largely based on analytical methods developed specifically for Gas Pixel Detectors in~\cite{bella03,Fab&Mul2014}, with subsequent refinements such as weighted event analyses~\cite{DiMarco2022}. The resulting approach, commonly referred to as the \emph{Method of Moments} (MoM), remains the standard reconstruction technique in this field, providing reliable estimates of both the emission angle and the impact point, and serving as a benchmark for alternative approaches. The increasing complexity of detector data has motivated not only the refinement of analytical methods, but also the exploration of machine learning approaches. However, the advantages and limitations of such methods compared to established techniques, particularly in terms of physically relevant observables derived from aggregated measurements, remain still unclear.}

These ML approaches have applied CNNs to photoelectron tracks by converting them into images. In the first work of this kind~\cite{kita19}, the authors framed angle reconstruction as a classification task, while those of~\cite{mor20} recast it as a regression problem. 
The study in~\cite{peir21} extends this approach with a deep ensemble of CNNs, but the conversion from the hexagonal pixel geometry of the detector to a square one still introduces systematic artifacts. 
When validated on real data~\cite{spie}, the method produces spurious modulation in unpolarized sources and distorts the reconstructed spectra. 
In~\cite{cib23}, the authors use CNNs only to reconstruct the impact point of the photoelectron, which is then fed to the MoM for angle reconstruction. 
The CNNs, therefore, do not predict the angle directly but indirectly improve angular reconstruction quality by providing more accurate impact point estimates. 
The authors of~\cite{li25} apply deep ensemble CNNs with layers adapted to the hexagonal pixel geometry of GPD data. 
They report a 15–30\% improvement in modulation factor compared to MoM; however, the evaluation is based on relatively small datasets, making the robustness of these gains uncertain. 
Finally,~\cite{Jiao2025} employs a simple CNN in a classification setup trained on an unusually large dataset of 20 million simulated tracks. 
Although they report improvements in modulation factor over MoM, low spurious modulation is achieved only through post hoc correction rather than through architectural design.

In all previous studies, CNNs have been used to process GPD data by converting the detector readouts into images. 
\rev{The conversion from hexagonal GPD tracks to square CNN images is typically achieved by mapping the hexagonal grid onto a Cartesian 2D array. Due to the staggered arrangement of hexagonal pixels, this transformation introduces ambiguities in the spatial alignment, particularly between even and odd rows. Different strategies have been proposed to address this issue, including duplicating representations to account for possible shifts, preserving row parity during preprocessing, or interpolating pixel values onto a square grid~\cite{peir21,cib23,Jiao2025}. However, all these approaches alter the native geometry of the detector, and the resulting track representation.}
Furthermore, photoelectron tracks are extremely sparse, often activating only a few pixels per event, which further complicates learning. Even the introduction of hexagonal convolutions in~\cite{li25} \rev{ did not fully overcome these challenges, suggesting that sparsity and geometric irregularity remain challenging aspects for CNN-based approaches.}
\rev{Moreover, previous works are typically trained and evaluated on datasets generated using  a source with a fixed energy (monochromatic) bins~\cite{kita19,cib23,Jiao2025} or with uniform emission in a certain energy band (flat energy  distributions~\cite{peir21,Jiao2025}, which provide a more controlled but less realistic setting compared to the spectra encountered in astrophysical observations.  In fact, astrophysical sources have emissions with more complex patterns; the more common ones are blackbody-like or power-laws.
In addition, these studies primarily assess performance in terms of global observables, such as the modulation factor, and rarely explicitly analyze the reconstruction accuracy at the level of individual events.}

In this work, we \rev{investigate} these limitations by modeling each photoelectron (PE) track as a graph, \rev{a representation that naturally preserves the native hexagonal structure and sparsity of the detector data}, and employing \emph{graph neural networks} (GNNs)~\cite{GNN19} that directly operate on the native hexagonal structure of the detector’s charge distribution to predict the initial emission angle, \rev{and compare their performance with the MoM, the standard approach, to investigate the trade-offs of using learning methods rather than the well-established analytic ones}. 
Training is performed exclusively on data from an unpolarized source, a deliberate design choice to decouple reconstruction from any latent polarization bias. 
This strategy suppresses spurious polarization artifacts that can arise when the reconstruction process unintentionally introduces polarization into intrinsically unpolarized data. 
Such artifacts are particularly problematic since most celestial sources exhibit polarization levels of only a few percent, meaning even small systematic modulations can significantly compromise measurement reliability. 
To further mitigate potential biases introduced by the neural network, we employ rotational data augmentation and ensemble selection. 
Finally, to our knowledge, this is the first work to train on tracks generated from an astrophysically motivated power-law photon spectrum, rather than on fixed monoenergetic bins or uniformly sampled energies. \rev{Such an energy distribution reflects those encountered in actual X-ray observations, making data more realistic, even if dominated by low-energy tracks, which are smaller, more difficult to reconstruct, and less informative. This makes the task for the GNN more challenging.}

GNNs have already shown strong performance on irregular and sparse domains in computer vision and the physical sciences~\cite{GNNs}.  
They provide a principled way to propagate geometric information through message passing on non-Euclidean structures such as meshes, point clouds, and detector graphs.  
Widely cited works illustrate how geometric deep learning unifies such architectures under symmetry and invariance principles~\cite{Bronstein2021, Battaglia2018}, and surveys confirm their success across domains including 3D vision, molecular modeling, and scientific instrumentation~\cite{Wu2020, Han2024}.  

In this context, the problem can be formally stated as follows: given the sparse detector readouts for a track recorded by the sensor on a hexagonal grid, the objective is to learn a function $f_\theta$ that predicts the photoelectron emission angle $\hat{\theta}$.
\rev{From a learning perspective, this corresponds to a regression task aimed at minimizing a per-event reconstruction error.
These per-event predictions are then aggregated to estimate the angular distribution of the signal. However, improvements at the local level do not trivially translate into better estimates of global observables derived from aggregated measurements. This separation can lead to counterintuitive outcomes, where improvements in per-event reconstruction error do not translate into, and may even degrade, the accuracy of physically relevant aggregate quantities such as the modulation factor. This is an important difference with respect to other CNNs where the final target was to maximize the modulation factor, also at cost of introducing bias in the response to unpolarized signals.}
\rev{Our work provides the following contributions:}
\begin{itemize}

\item\rev{ \textbf{Local vs global reconstruction trade-off.}
We analyze the relationship between per-event reconstruction accuracy and the estimation of global observables, showing that improvements in per-track precision do not necessarily translate into better reconstruction of the angular distribution. This highlights a \revv{possible} limitation in learning-based approaches to this problem.}
\item \rev{
\textbf{Geometry-aware learning framework.}
We develop a GNN-based reconstruction pipeline, combining graph-based data representation, rotational data augmentation, modulation-aware model selection, and ensemble aggregation to mitigate spurious polarization signals. This model operates directly on the native detector geometry, providing a suitable representation for irregular and sparse data.}

\item \textbf{Astrophysically motivated training sample.}
We train and evaluate the model on tracks generated from a power-law photon spectrum, rather than on monoenergetic or uniformly sampled energies, resulting in a more realistic and intrinsically more challenging reconstruction setting.

\end{itemize} 


\rev{These elements define a reconstruction framework designed to limit spurious polarization effects, which we use to analyze the interplay between local reconstruction accuracy and global performance in realistic polarimetric scenarios.}



The paper is organized as follows: In the next section, we explain the \rev{track angle} reconstruction problem from an astrophysical perspective. In section 3, we state the considered machine learning problem, the use of GNNs, their suitability to the problem, and the elements that compose the neural network. In section 4, we report detailed results with a comparison with the geometric method generally used in the literature. Finally, in section 5, we report the conclusions. 

\section{Astrophysical Motivation and Characteristics of GPD Track Data}\label{sec: the sensor}

X-ray polarimetry based on the photoelectric effect~\cite{Heitler1954} is currently the most effective method to measure polarization in the 2--10~keV energy band, which is considered the classical range for X-ray polarimetric studies. In this band, X-ray sources are sufficiently bright, and focusing optics, such as grazing-incidence mirrors, are highly efficient, making it the preferred range for most proposed space missions.

One of the most prominent recent efforts to exploit the advantages of photoelectric polarimetry in this energy range is the Imaging X-ray Polarimetry Explorer (IXPE, \cite{Weisskopf2022,Soffitta2021}), proposed in January 2017 and launched in December 2021. IXPE consists of three identical telescopes, each comprising a mirror module and a focal plane Detector Unit (DU). Each DU \cite{Soffitta2021}) hosts a GPD \cite{Baldini2021}, a technology developed over more than 25 years of continuous advancement, from the first one-dimensional detector \cite{Soffitta2001} to fully two-dimensional devices optimized for sensitive satellite-borne polarimetry \cite{Costa2001,Bell2006,Bell2007,Baldini2021}.   
IXPE has detected X-ray polarization from several dozen celestial sources, enabling measurements of the polarization degree and angle from compact objects such as isolated pulsars, neutron stars, black hole binaries, and active galactic nuclei (AGN). In addition, it has mapped the polarization of extended sources, including pulsar wind nebulae, supernova remnants, molecular clouds near the Galactic Center, and the lobes of interacting jets from black hole binaries such as SS~433, providing insights into their interaction with the surrounding environment.

The Gas Pixel Detector (GPD) consists of a gas cell with a thin beryllium window, a drift region filled with a polarimetry-optimized gas mixture (dimethyl ether, DME), and a Gas Electron Multiplier (GEM) that amplifies the tracks produced by photoelectrons. The amplified charge is collected by an Application Specific Integrated Circuit (ASIC) developed specifically for this purpose~\cite{Bell2006}. This ASIC features 105{,}600 hexagonal pixels with a 50~$\mu$m pitch and outputs the analog signals corresponding to the charge collected by each pixel (which is directly related to the photon energy and referred to as ``energy'' in the following sections). It also provides the digital coordinates of the Region of Interest (RoI), defined by local triggers that identify tracks and include a padding of 20 pixels. In addition, a trigger signal is generated for each RoI to enable time-tagging of the event.
The analog signals are digitized by the back-end electronics and transmitted to the Detector Service Unit, which packages the data and stores it in the onboard memory.

Unlike conventional Charge-Coupled Devices (CCDs), the Gas Pixel Detector (GPD) is self-triggered and reads out only a subset of the total pixel array. This significantly reduces dead time to a manageable level. The output consists of a stream of pixel values with corresponding coordinates, arranged in a hexagonal pattern.

In contrast to conventional pixellated X-ray detectors, the GPD, through its ASIC-based pixel pattern, resolves the photoelectron track. This allows the emission direction to be determined with high precision from the impact point by applying a single algorithm. The algorithm calculates the first three moments of the track, effectively distinguishing the impact point from the opposite end of the track, which appears as a skein marking the end of the electron's path (see, e.g., \cite{DiMarco2022}).

The size of the photoelectron track depends on the photon energy: at higher energies, it can extend over hundreds of pixels, while at lower energies, where the number of detected counts is significantly higher because of the typical spectra of X-ray sources, the mirror effective area, and detector efficiency, the track may cover only a handful of pixels. In this regime, hexagonal pixels offer a distinct advantage over square pixels. They reduce the systematic effects associated with the preferential directions along the sides of square pixels compared to those along their diagonals. This minimization of the directional bias is the main reason for adopting a hexagonal pixel pattern in the GPD design. 
 
These detector characteristics motivate the need for learning architectures that can natively operate on irregular geometries and sparse event data—requirements well matched by graph neural networks, as explored in the following section.

\section{Graph Neural Networks for Track Angle Reconstruction}

While GNNs have been widely explored in other scientific domains, their application to gas pixel detector data introduces distinct challenges: the need to preserve hexagonal topology, handle extreme sparsity, and control angular modulation at sub-percent levels. Our contribution lies not merely in adopting a GNN architecture but in designing a learning protocol that is explicitly aware of the physical symmetry constraints of the detector and the statistical requirements of polarization analysis.

We focus on training and testing neural networks to determine the emission direction of PEs, with the goal of minimizing any systematic modulation in the angular distribution produced by the model in response to unpolarized X-ray sources. This task is formulated as a regression problem, where each track is assigned a predicted value in the range $[-\pi, \pi]$.

\begin{figure}[h]
\centering
\includegraphics[scale=0.4]{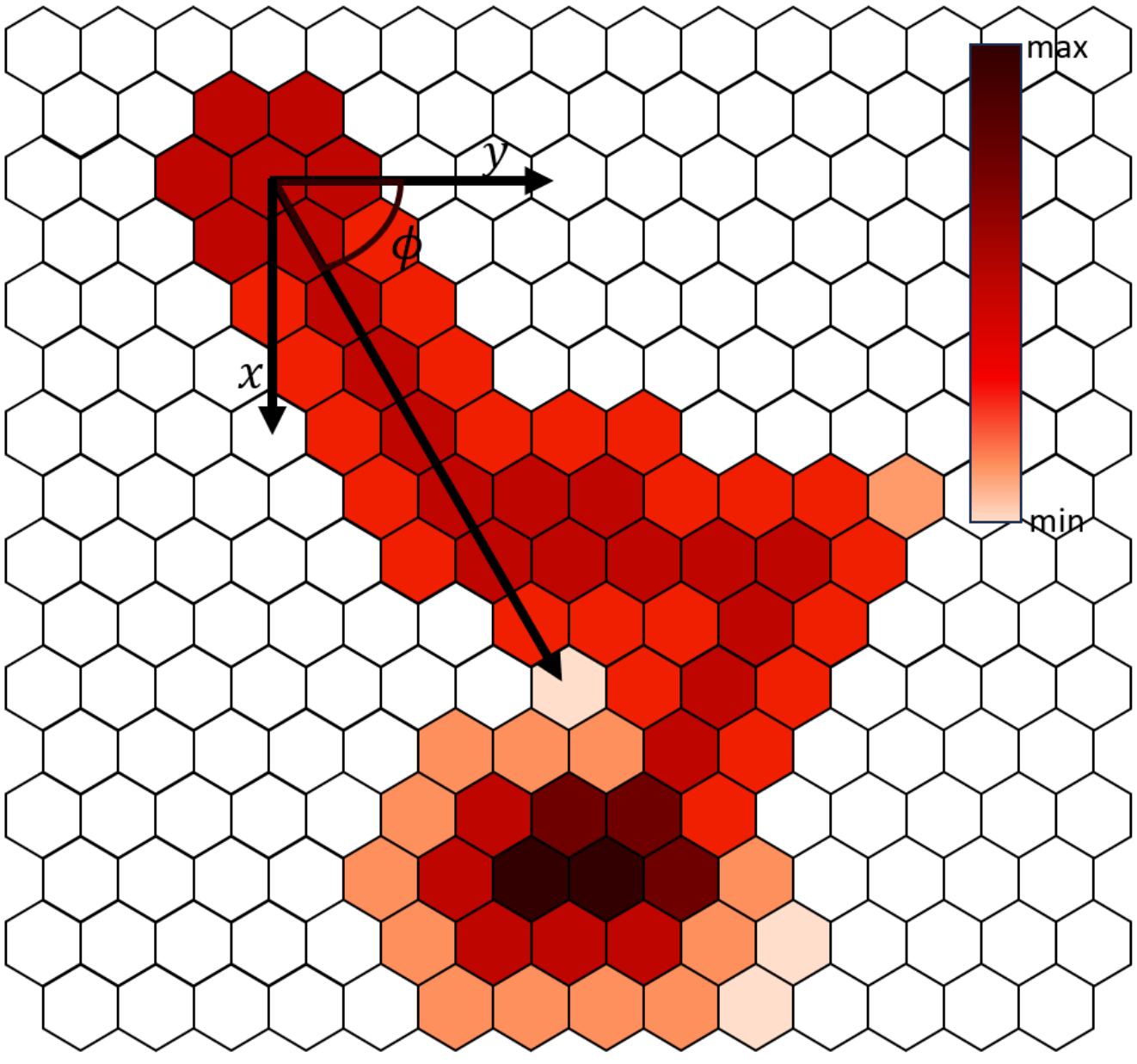}
\caption{A track imaged by the GPD. Each hexagonal pixel represents a detector sensor, and its color encodes the collected charge (number of electrons) at that pixel. The color scale is normalized on a per-track basis, highlighting relative variations within the track. \revv{The long diagonal arrow indicates the emission direction of the photoelectron. The angle $\phi$ is measured with respect to the detector $y$-axis; for a polarized photon beam, the electric field (polarization vector) oscillates along this direction.}}
\label{fig:track}
\end{figure}

As discussed in the Introduction, CNN-based methods require converting hexagonal GPD tracks, as illustrated in Figure~\ref{fig:track}, into square images, which introduces distortions and fails to take advantage of the sparsity of the data.
\rev{Even though the photoelectron track is inherently three-dimensional, the detector measures its projection onto the readout plane where the sensors are located. Consequently, the emission angle considered in this work corresponds to the azimuthal angle of this projected track.}

\begin{figure*}[h]
\centering
\includegraphics[scale=0.22]{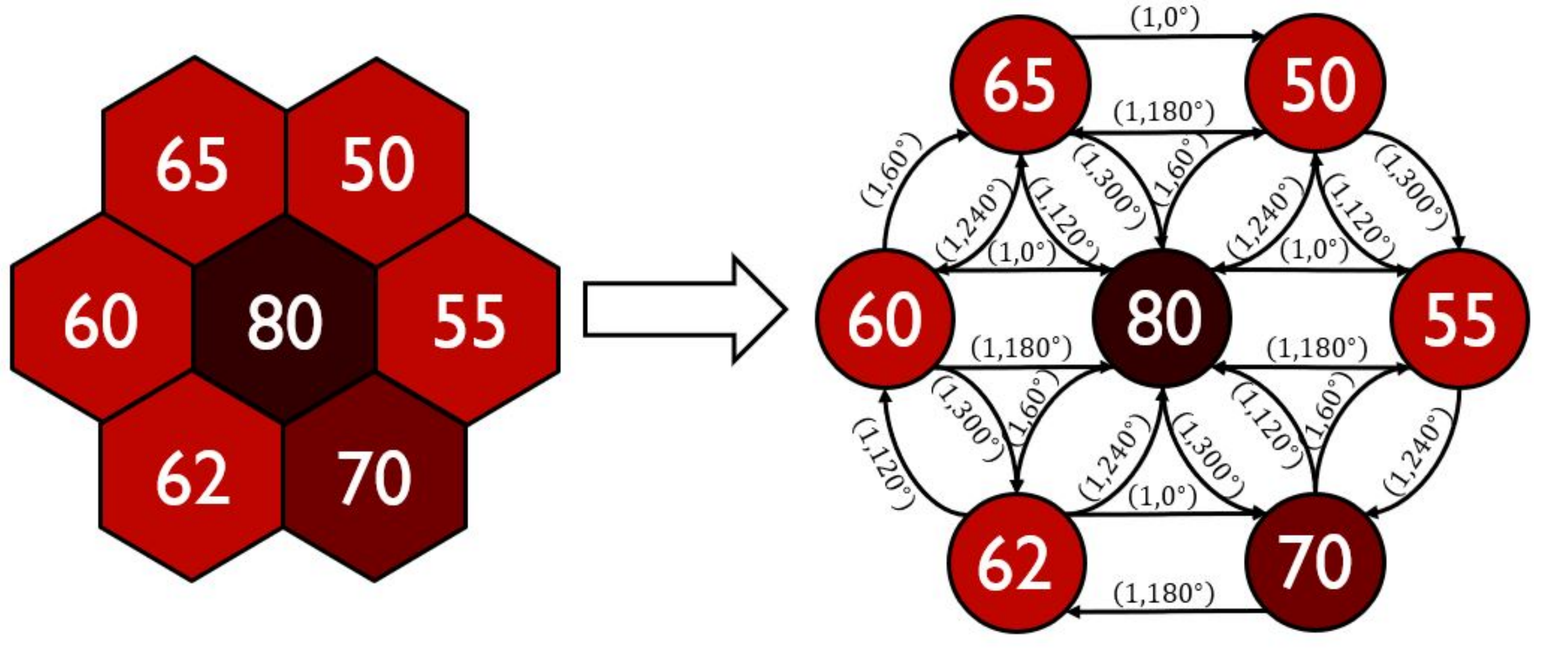}
\caption{\rev{Conversion of a hexagonal GPD cell into a graph. Each pixel becomes a node with the detected energy as a node attribute. Edges connect adjacent nodes and encode relative polar coordinates. For each pair of neighboring nodes, two directed edges are included, with arrows indicating direction and labels the corresponding polar coordinates.}}

\label{fig: conv}
\end{figure*}
To avoid such limitations and to provide a data representation that more closely reflects the original tracks generated by the GPD, we represent each track as a graph and employ GNNs to reconstruct and predict the impact angle of the photoelectron responsible for the track.

The conversion from track to graph is straightforward and is illustrated in Figure~\ref{fig: conv}. Each active pixel in the track corresponds to a node in the graph, and the charge deposited on that pixel is retained as a node-level attribute. Each edge carries the pseudo-polar coordinates between the connected nodes as attributes, providing relative positional information that describes how one node is situated with respect to its neighbors. The global geometry of the track is preserved through these local, edge-wise geometric relations, which are propagated across the graph via node adjacency.

We consider edges only between immediately adjacent nodes. This restriction is necessary because the number of edges grows quadratically with the number of nodes, leading to excessively large graphs and high memory usage. As a result, the first polar coordinate, representing radial distance, is always equal to one, since the hexagonal pixels are equidistant; this coordinate can therefore be omitted during training. The second polar coordinate corresponds to one of six discrete angles: $\{0^\circ, 60^\circ, 120^\circ, 180^\circ, 240^\circ, 300^\circ\}$, reflecting the hexagonal symmetry of the detector.

This representation enables a sparse encoding of the graph, in which only the nodes activated by the energy deposited by the photoelectron are input to the neural network. Moreover, the hexagonal organization of the pixels is naturally preserved and managed by the graph structure through the geometric information encoded in the edges.

Another advantage of graph representation is that longer tracks are never truncated. In contrast, images input to CNNs must have a fixed size, which often necessitates cropping longer tracks to fit within the predetermined image dimensions dictated by the input layer of the network. For example, while a short track may easily fit within a $30 \times 30$ image, a longer track containing more information might require a $60 \times 60$ image. However, increasing the size of the image leads to increased computational complexity and longer training times.
In the graph-based approach, the number of nodes can vary from graph to graph, eliminating the need to discard information. In our experiments, the number of nodes ranges from 60 to 600, demonstrating that tracks of varying lengths can be processed effectively using the proposed methodology.

\subsection{Dataset generation and Angular Modulation Control through Data Augmentation}
\begin{figure}[h]
\centering
\includegraphics[scale=0.55]{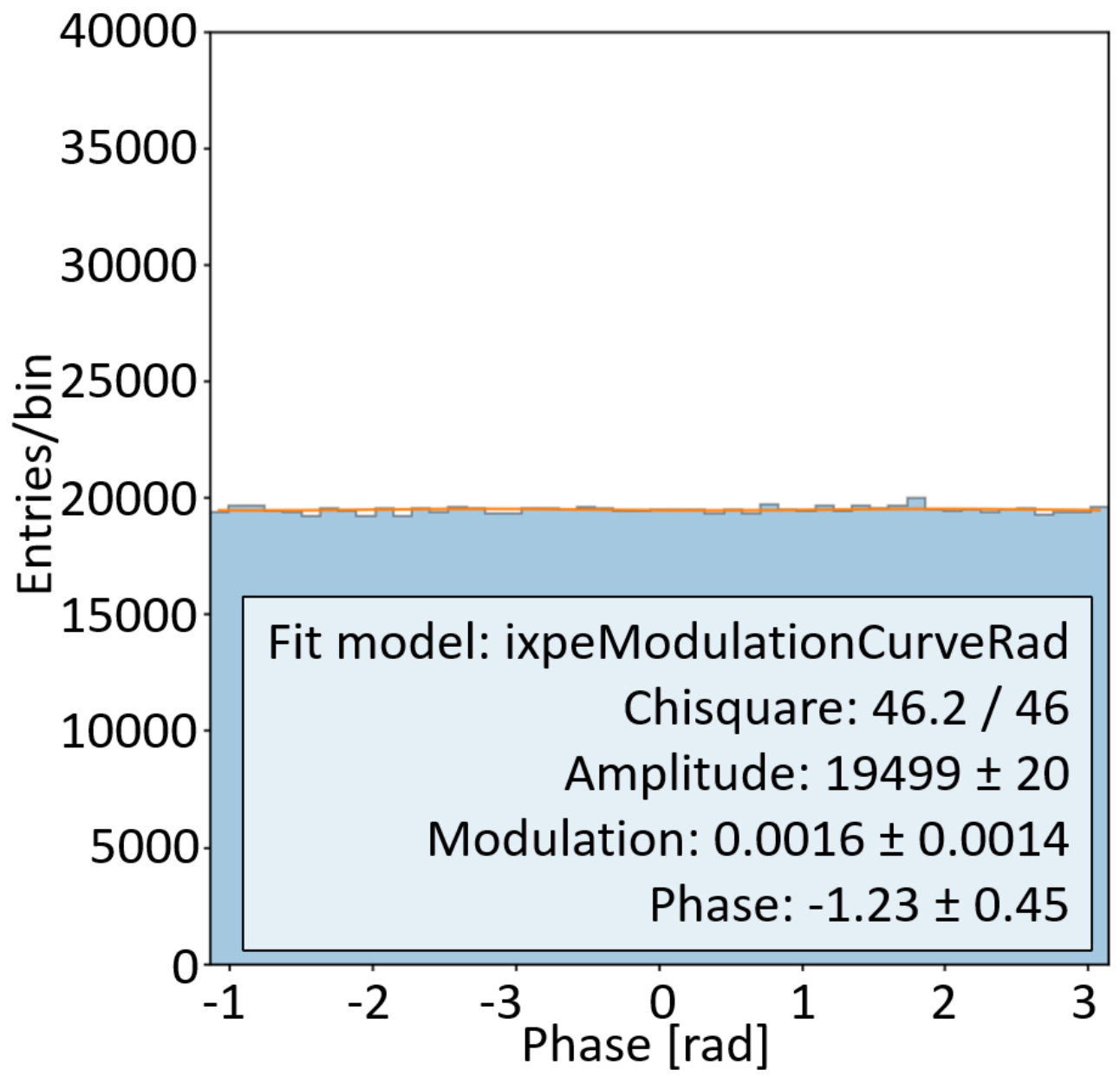}
\caption{Angular distribution of \rev{reconstructed} emission angles for simulated unpolarized tracks, showing near-zero modulation over $[-\pi, \pi]$. \rev{The orange curve represents the best-fit modulation function $\mathcal{M}_\phi$.  The Amplitude, Modulation and Phase are referred to the function $\mathcal{M}_\phi$ defined in Equation \eqref{eq:mod}, while the Chi-square, defined in  Appendix \ref{app:phys}, refers to its fit.}}
\label{fig: flatex}
\end{figure}

The tracks used for training are generated using Monte Carlo simulation software \textsc{IXPESIM}~\cite{bella03,bal22}, developed to accurately simulate the response of the GPD. The datasets for training and validation are produced using an unpolarized light beam. The tracks generated under this condition exhibit a uniform distribution of emission angles over the interval $[-\pi, \pi]$, as illustrated in Figure~\ref{fig: flatex}. 
\rev{This is possible for data provided by a Monte Carlo code; in fact in such a case we know the true angle of emission of the photoelectron and we can compare this one with the GNN prediction. The choice to train the model exclusively on tracks generated by an unpolarized light beam is deliberate, as it prevents the network from learning spurious correlations associated with specific polarization patterns, which could introduce systematic biases (residual modulation). If polarized tracks were included in the training, the model could incorrectly interpret noise as signal or distort the reconstructed polarization pattern.}
Training, validation, and test sets all have a dimension of one million tracks each.

Regarding the energy distribution used to generate the tracks, we adopt a power-law photon spectrum with an exponential index of $-2$, producing a set of tracks with continuously varying energies. 
\rev{The power-law spectrum with an exponent index of $-2$ well describes the Crab Nebula, that is considered an archetypal X-ray source due to its brightness, stability, and well-characterized synchrotron spectrum \cite{buc23} (also check Appendix \ref{app:phys} for more details).}
This spectrum reflects a realistic astrophysical scenario and makes the prediction task more challenging, as the resulting excess of low-energy tracks, which are shorter and less informative, amplifies the difficulty of emission angle reconstruction. Although the IXPE energy band is between 2 and 8 keV, we deliberately train the entire energy spectrum to evaluate how the network generalizes across all energy levels, including those outside the operational band.
The IXPE gas pressure is set to 644 mbar in the simulations.

We train the neural network on unpolarized data to ensure that it learns from inputs that do not encode any intrinsic angular information and therefore exhibit no periodic variation in the angular domain, such as sinusoidal modulation. This strategy prevents the network from developing bias toward any particular polarization signal. As a result, the level of bias in a neural network can be quantified through the amplitude of the modulation measured in this condition,  $\mu$, defined as:
\begin{equation}
\mu = \frac{\max(\mathcal{M}_\phi) - \min(\mathcal{M}_\phi)}{\max(\mathcal{M}_\phi) + \min(\mathcal{M}_\phi)} 
\end{equation}
where $\mathcal{M}_\phi$ is the angular distribution of the reconstructed photoelectron angles \rev{which is called Modulation Curve. $\phi$ is the so called azimuthal angle and represents the angle, in the plane perpendicular to the photon direction, between the projection of the reconstructed emission direction and some reference axis of the instrument; the \textit{Phase} $\phi_0$ indicates the direction of polarization.}
\rev{ In fact we use as fitting function: 
\begin{equation}\label{eq:mod}
\mathcal{M}_\phi = Amplitude \times [1 + Modulation \times \cos(2(\phi-\phi_0)]
\end{equation}}
\rev{The \textit{Modulation}, in the case of 100\% polarized radiation is called \textit{Modulation Factor}.  In this paper we call $\mu$ the generic modulation albeit it usually indicates the  \textit{Modulation Factor}.} 
 
\rev{In the figures reporting angular distributions, such function is shown as the orange curve while in the figure legend \textit{Amplitude} and \textit{Modulation} are those defined in Eq. \eqref{eq:mod}}. Since the angular distribution of the original unpolarized emission typically exhibits a modulation of approximately below 0.2\% (see Figure~\ref{fig: flatex}), we consider an amplitude modulation factor below 2\%, with a statistical error under 0.02\%, to be acceptable for unpolarized datasets predicted by the neural network. The modulation amplitude values below 2\% for unpolarized sources are also reported by the classical geometrical reconstruction methods used as benchmarks against machine learning approaches, as shown in~\cite{spie, cib23}.
Furthermore, in our empirical experiments, certain artifacts introduced by some networks in the angular distribution become apparent only when datasets of a size of at least $5 \times 10^5$ samples or larger are analyzed. Consequently, the modulation factor should be evaluated on datasets of such size  to ensure statistical reliability. 

If these artifacts are too pronounced, the model may falsely detect polarization signals when none are present or distort the true polarization pattern. Therefore, maintaining a sufficiently low modulation factor on unpolarized datasets is a critical property of any model, as it indicates the absence of systematic distortions.

However, achieving low modulation is not trivial and requires careful consideration. For example, in~\cite{kita19}, the authors implement a division-and-rotation strategy to mitigate systematic irregular modulations in neural network predictions, successfully reducing the modulation in some cases to below 1\%. They also incorporate a chi-square term into the loss function to encourage flatness in the predicted angular distribution for unpolarized signals.
Although this is a useful strategy, it introduces a notable drawback: the chi-square loss is computed only over the batch of tracks currently fed to the network during training. As a result, even if each batch contains hundreds of samples, the chi-square statistic may deviate significantly from its true value over the full angular distribution, potentially leading to suboptimal or unstable modulation suppression.

\begin{figure*}[h]
\centering
\includegraphics[scale=0.2]{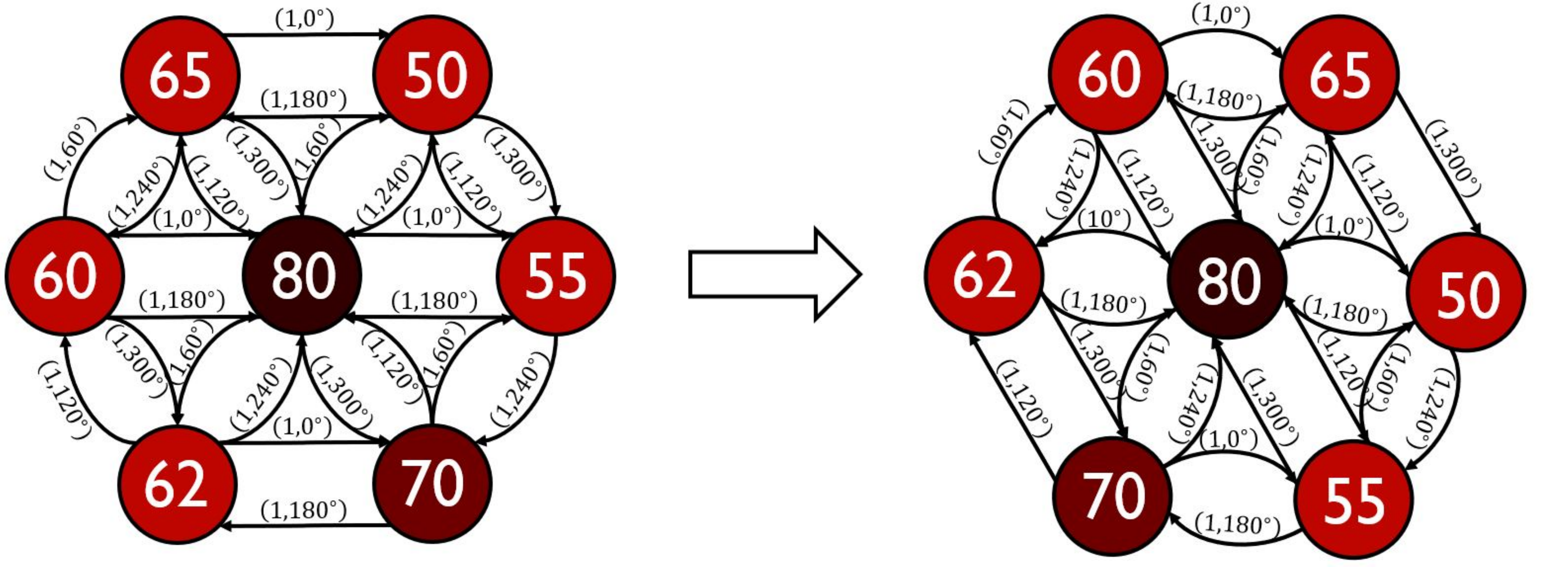}
\caption{Example of a $60^\circ$ clockwise rotation of a graph. The node attributes remain unchanged, while only the edge attribute corresponding to the second polar coordinate is decreased by $60^\circ$ to reflect the rotation.}
\label{fig: rot}
\end{figure*}

For this reason, in our experiments, we adopt a different strategy based on data augmentation through track rotation, as previously employed in~\cite{peir21,spie}. For each original track, we generate five additional rotated versions applying rotations of $\{60^\circ, 120^\circ, 180^\circ, 240^\circ, 300^\circ\}$. 
Notably, because of the graph-based representation of the tracks, these rotations can be implemented in a straightforward and exact manner. The node attributes and the first polar coordinate remain unchanged, while only the second polar coordinate requires adjustment. For example, a $60^\circ$ clockwise rotation is applied by subtracting $60^\circ$ from the second polar coordinate, as illustrated in Figure~\ref{fig: rot}. Furthermore, the graph representation preserves rotational equivariance more naturally, making this form of augmentation particularly effective.
This graph-based rotation avoids the potential distortions that may arise when rotating image-based representations. In particular, converting a track into a square-pixel image and then applying a $60^\circ$ rotation, as done in~\cite{peir21}, can introduce interpolation artifacts and geometric inconsistencies, issues that are entirely circumvented by directly rotating the graph. In addition, image-based operations may introduce additional artifacts and lead to an increase in spurious modulation.

During training, the rotated graphs are randomly fed into the network to avoid potential biases that could arise from presenting consecutive rotations of the same track. 

\begin{figure*}[h]
\centering
\includegraphics[scale=0.5]{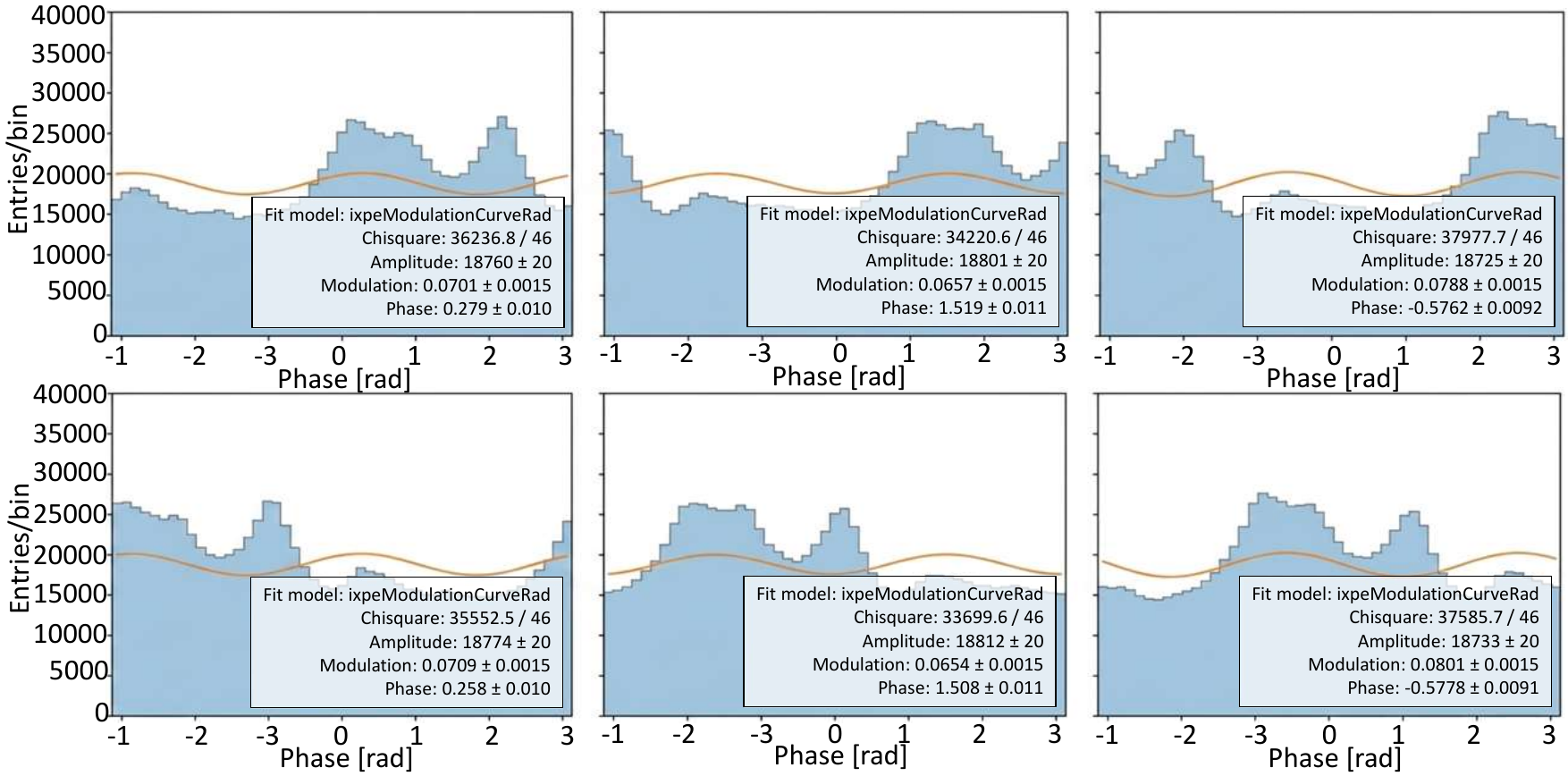}
\caption{Predicted angle distributions from the neural network for six rotated versions of a dataset originally composed of one million unpolarized tracks. Each distribution exhibits angular bias introduced by the model. \rev{The orange curve represents the best-fit modulation and it is relative to the distribution in its respective sub-figure.}  Notably, the peaks and valleys are offset by $60^\circ$ between successive rotations, illustrating how rotational artifacts shift predictably across orientations.}
\label{fig: augs}
\end{figure*}

During inference on the test set, once the network has been trained, the final predicted angle for a track is computed as follows:
\begin{enumerate}
    \item Collect the six predicted angles corresponding to the six rotated versions of the track at $\{0^\circ, 60^\circ, 120^\circ, 180^\circ,$ $ 240^\circ, 300^\circ\}$;
    \item Rotate each of these predicted angles back to the original track orientation;
    \item Compute the final predicted angle as the average of the six corrected values.
\end{enumerate}
This strategy effectively mitigates systematic artifacts introduced by the neural network in the angular distribution. If the model tends to introduce orientation-dependent biases, averaging predictions across the six rotationally shifted versions helps to cancel them out.
This effect is illustrated in Figure~\ref{fig: augs}, which shows the predicted angle distributions produced by the neural network for one million unpolarized tracks, evaluated separately for each of the six rotations. Although none of the individual distributions are flat—each exhibits peaks and valleys introduced by the model—the bias patterns are consistently shifted by $60^\circ$ between successive sub-figures. As a result, when predictions are averaged across these rotations, the individual biases largely cancel out, yielding a significantly flatter overall angular distribution, as discussed in the results section.

The distributions shown in Figure~\ref{fig: augs}, when considered individually, illustrate the angular bias introduced by the network in the absence of rotational data enhancement, along with artifacts that affect the entire predicted angle distribution. Without this augmentation, such distortions persist in the final output, leading to a markedly non-uniform angular distribution. 
On polarized datasets, these same artifacts can interfere with the underlying sinusoidal modulation signal, thereby reducing the fidelity of the measured modulation factor. Consequently, the effects observed in Figure~\ref{fig: augs} directly reflect the behavior of the model when rotational augmentation is not used.

\subsection{The GNN architecture}
At their core, GNNs operate by propagating and aggregating information across a graph through iterative updates of each node's representation, based on the features of its neighbors and the structure of the connecting edges. This message-passing framework enables GNNs to model complex dependencies and interactions in sparse, relational data, effectively capturing both local and global structural patterns.

\begin{figure*}[h]
\centering
\includegraphics[scale=0.23]{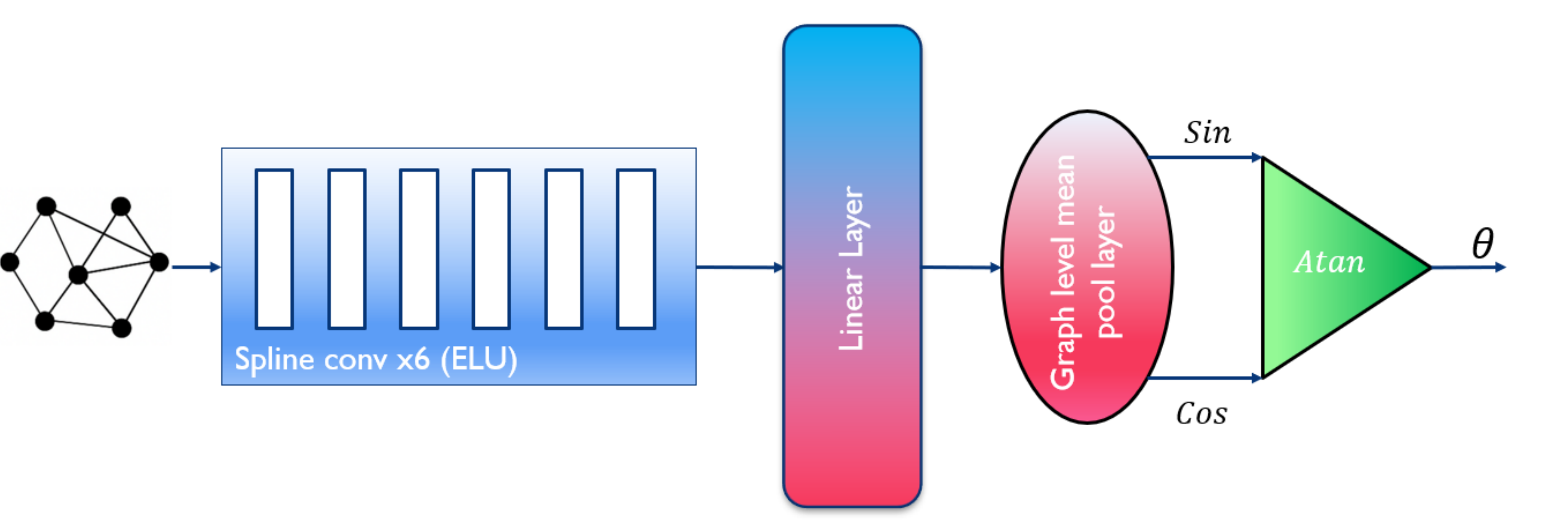}
\caption{GNN architecture for photoelectron angle reconstruction: six spline convolution layers (ELU activation), a linear transition layer, a graph-level mean pooling layer, and an \texttt{atan2} operation that combines the predicted sine and cosine values into the reconstructed angle $\theta$.}
\label{fig: augs1}
\end{figure*}

We construct the architecture of our network using the standard spline convolutional layer introduced in~\cite{fey18} as shown in Figure \ref{fig: augs1}. The implementation is based on PyTorch~\cite{pytorch}, specifically utilizing the PyTorch Geometric library~\cite{GNN19}.
Spline convolution is selected after evaluating several standard GNN variants during the initial development phase. Although relatively simple, this architecture consistently yields stable training and low modulation, primarily due to its ability to handle continuous edge attributes, such as polar coordinates between nodes. Since our primary objective is to reduce systematic angular bias rather than optimize network complexity, we adopted this effective and computationally efficient design throughout our experiments. We emphasize that while the architecture itself is intentionally simple, the novelty of our approach lies in the graph-based data representation and training strategy designed to suppress spurious modulation.

The network consists of six convolutional spline layers, followed by a linear layer and a graph-level prediction head. Each convolutional layer uses a kernel size of five, and an exponential linear unit (ELU) activation function is applied after each convolutional and linear layer.
Following the linear layer, a graph-level pool layer is applied to perform the prediction task on entire tracks, rather than on individual nodes or edges. For this layer, we use a global mean pooling operation, which aggregates node features and outputs two values corresponding to the sine and cosine of the predicted angle. These two quantities are then passed to an arc-tangent function, ensuring that the final network output is a real number in the interval $[-\pi, \pi]$.
For optimization, we use the Adam algorithm~\cite{adam} with an initial learning rate of $10^{-3}$ .

Regarding the loss function, the predicted and true angles are not directly compared in radians. This is because the angles lie within the interval $[-\pi, \pi]$, and two values that are numerically distant in this range may, in fact, represent the same or very similar directions. For example, directly comparing $-\pi$ and $\pi$ would yield an error of $2\pi \approx 6.283$, even though both correspond to the same angle of $180^\circ$.
To address this, each angle is first converted into its sine and cosine components, and the corresponding components are compared using PyTorch’s Smooth L1 loss function. For a single component (e.g., the sine), the loss is defined as:
\begin{equation}
    \begin{small}
\text{SmoothL1}(x, y) =
\begin{cases}
0.5 \cdot (f(x) - y)^2 / \beta & \text{if } |f(x) - y| < \beta, \\
|f(x) - y| - 0.5 \cdot \beta   & \text{otherwise},
\end{cases}
\end{small}
\end{equation}
where \( x \) is the input track, \( f(x) \) is the predicted sine (or cosine) component, and \( y \) is the corresponding ground truth value. We emphasize that when training with rotated tracks, the emission angle associated with them \( y \) is also rotated accordingly. The total loss over the dataset is then computed by summing the Smooth L1 losses for the sine and cosine components across all training samples.

The sensitivity parameter is set to $\beta = 0.5$. This value is chosen because small discrepancies between the reconstructed and true angles are acceptable in our application, where the objective is to recover the general orientation of the light rays that generate the photoelectron tracks, rather than achieving high precision for each individual angle. The choice $\beta = 0.5$ ensures that angular differences below $30^\circ$ are penalized less severely, aligning the loss function with the physical tolerances of the problem.

Training is performed using 16 hidden channels per layer for 100 epochs, with a batch size of 2000 elements. For training, validation and test sets, we initially generate one million tracks each, which increases to six million per dataset after applying data augmentation through rotations.
During each epoch’s validation phase, we compute the modulation amplitude throughout the validation set. This estimate is reliable, as it is calculated only after predicting the angles of all tracks in the data set. However, during training, we observe that the modulation amplitude does not correlate directly with the validation loss. In other words, two epochs with similar validation loss values can yield substantially different modulation levels.
To address this, at the end of training, we first select the epoch with the lowest validation loss, then identify all epochs with validation loss within 1\% of this minimum, and finally choose among these the epoch with the lowest modulation. This strategy results in final models that maintain nearly the same prediction accuracy while reducing the modulation by more than two-thirds.

This training procedure is repeated ten times with different random seeds, resulting in an ensemble of ten distinct neural networks. Predictions from all possible non-empty subsets of the ensemble (i.e., $2^{10} - 1 = 1023$ combinations) are then aggregated using angular averaging. The subset that produces the lowest modulation in the validation set is selected for the final inference. It should be noted that, for all combinations of ensembles, the maximum validation loss differs from the minimum by only about 1\%, while the maximum aggregated modulation is more than three times higher than the lowest observed value. This shows that the ensemble approach effectively reduces modulation without compromising prediction accuracy.

The combination of techniques presented—graph-based data representation, rotational augmentation, careful selection of the training epoch based on both loss and modulation, and ensemble aggregation—collectively contributes to significantly reducing the modulation introduced by the model on unpolarized datasets. While each method offers partial mitigation on its own, their synergy proves especially effective. Rotational augmentation ensures that directional biases are evenly distributed and cancel out during inference. The graph-based representation avoids distortions commonly introduced by image-based rotations, thereby preserving angular integrity. Model selection based on modulation, rather than loss alone, helps prevent the choice of solutions that are numerically accurate but systematically biased. Finally, ensemble averaging further suppresses residual artifacts by leveraging the diverse inductive biases of independently trained networks.

\section{Results and Evaluation on GPD Track Data}
In this section, we report the results of our proposed methodology on three main datasets:
\begin{enumerate}
    \item An unpolarized dataset without signal, of the same type used to train the ensemble of networks;
    \item A 100\% polarized dataset characterized by a cosine-square angular distribution, peaking at $\theta = 0$ and $\theta = \pi$, and with a modulation equal to one;
    \item A second 100\% polarized dataset characterized by a cosine-square angular distribution shifted by $90^\circ$, peaking at $\theta = -\pi/2$ and $\theta = \pi/2$, also with a modulation equal to one.
\end{enumerate}

\rev{We adopt the cosine-square angular distribution for polarized sources, as it is the theoretically expected emission pattern for photoelectrons in the GPD.} Each dataset consists of one million samples to reveal possible artifacts that may emerge during the reconstruction of large datasets. All datasets are generated using IXPESIM~\cite{bella03,bal22} with a power-law photon spectrum of index $-2$. 
\rev{These datasets are used to evaluate both the reconstruction of global observables, such as the modulation of the angular distribution, and the accuracy of per-event angle predictions, enabling a direct comparison between these two levels of description.}

\rev{As previously stated, the primary benchmark method in our analysis is the Method of Moments (MoM)~\cite{bella03,Fab&Mul2014,DiMarco2022}, which remains among the standard approaches for angle reconstruction in IXPE data analysis.}
We also reference and compare with the results from recent studies that employ neural networks to address similar problems in astrophysics.

\subsection{Modulation Results}
In this section, we focus on the modulation of the angle distribution for the three considered datasets.

\begin{figure*}[h]
\centering
\includegraphics[scale=0.5]{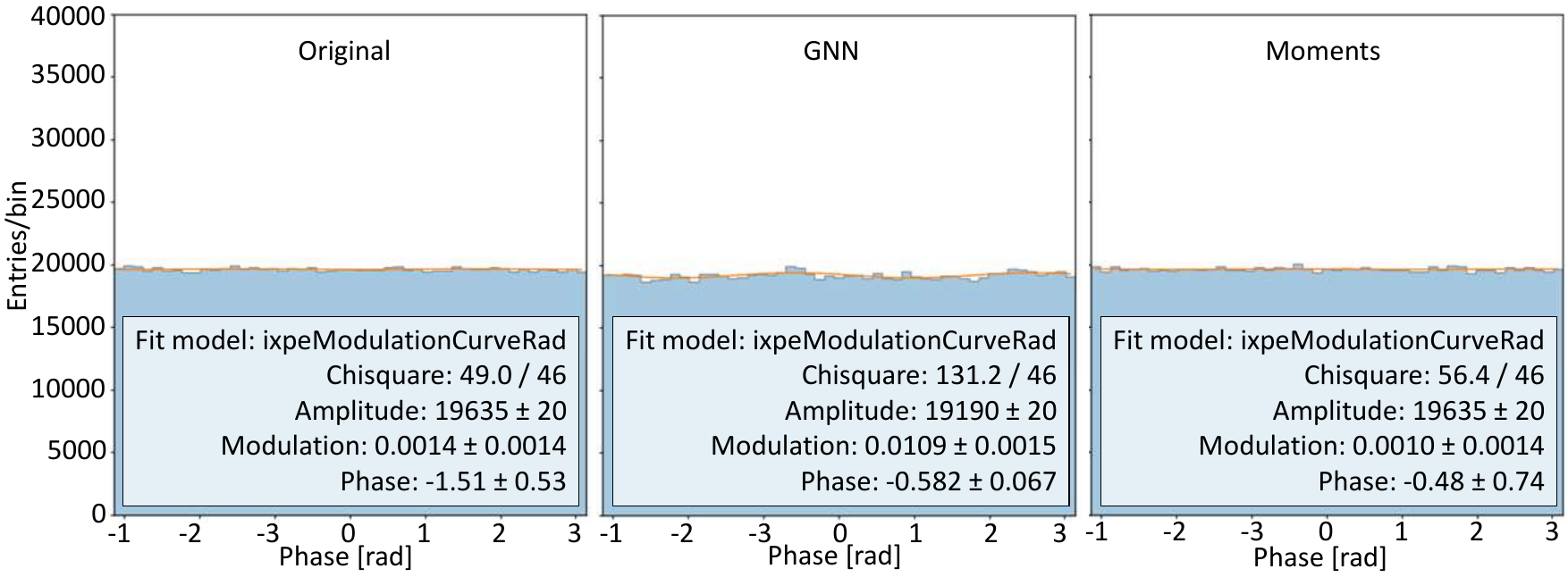}
\caption{Emission angle distributions for unpolarized data with a power-law spectrum over a million tracks: original angles (left), GNN reconstruction (center), and method of moment reconstruction (right). }
\label{fig: flat}
\end{figure*}
In Figure~\ref{fig: flat}, we report the angular distribution of the original emission source, along with the distributions reconstructed by the GNN and the MoM. 
The modulation of the original emission angle distribution is around 0.14\%, consistent with zero within statistical fluctuations, as expected. 
The GNN reconstruction exhibits a residual modulation of $\mu_{\text{GNN}} \simeq 1\%$, while the MoM yields $\mu_{\text{MoM}} \simeq 0.1\%$, which is effectively consistent with no polarization. \rev{These results show that, while the GNN is able to control spurious modulation at the percent level, the MoM on the full dataset remains significantly more effective in not introducing residual modulation, which is the primary requirement for unpolarized data. This highlights a clear difference in performance between the two approaches in terms of systematic bias.}

\begin{table}[h!]
\centering
\caption{Residual modulation on unpolarized datasets for different reconstruction methods.}
\renewcommand{\arraystretch}{1.15}
\begin{tabular}{p{1.cm} p{2.8cm} p{3cm} p{1.4cm} p{4.5cm}}
\hline
\textbf{Method / Ref.} &
\textbf{Data type test/ size} &
\textbf{Training setup} &
\textbf{Residual modulation (\%)} &
\textbf{Notes / caveats} \\
\hline
\textbf{GNN} &
1M simulated; power-law spectrum &
1M tracks; power-law spectrum; rotational aug.; ensemble avg. &
$\sim$1.0 &
No post-proc.; no harmonic artifacts \\
\textbf{MoM} &
\rev{ No neural network}, analytical solution &
\rev{ No training, but only} geometric reconstruction &
$\sim$0.1 &
Baseline; minimal modulation \\
~\cite{peir21} &
0.35M simulated; Flat spectrum 1–9 keV &
3.15M tracks; Flat 1–9 keV spectrum; CNN ensemble on square images; Unpolarized training; rotational aug.; 
&
\rev{$[0.1,0.7]$} &
No post-proc.; excluded scattered tracks; artifacts on real data \\

~\cite{spie} &
Measured from IXPE data &
Same CNN as~\cite{peir21} &
\rev{$[1,2]$} &
No post-proc.; spurious $n=6,12$ harmonics. Note this is a validation, not a training the results of this work is the validation of \cite{peir21}  \\

~\cite{cib23} &
$>$2.5M simulated; Monochromatic 2–8 keV spectrum, in 0.5 keV steps; 1.5M measured from astrophysical sources & 2M tracks; Flat 1–9 keV spectrum;
CNN for impact point + MoM for angle &
$\sim$0.1 &
Low modulation inherited from MoM \\

~\cite{li25} &
$\approx$$50k$ Simulated; Flat spectrum 1–9 keV &
$\approx$$800k$ Simulated; Flat spectrum 1–9 keV; Hex-CNN classification &
\rev{$[0,2]$} &
Limited data in test, not statistically robust; MoM baseline Modulation around $\sim$1–2\% \\

~\cite{Jiao2025} &
3.2M simulated with Discrete mono-energetic spectrum at 2.62, 3.74, 4.5, 5.9 keV; 2M  measured from astrophysical sources &
20M tracks; Discrete mono-energetic spectrum at 2.62, 3.74, 4.5, 5.9 keV; CNN classification &
\rev{$[0.2,0.5]$, after post-hoc cal.} &
Post-hoc calibration; uncorrected higher modulation \\ 
\hline
\end{tabular}
\label{tab:comparison}
\end{table}

Table~\ref{tab:comparison} summarizes the residual modulation reported for unpolarized datasets across recent approaches.
\rev{Unlike previous works, our model is trained and tested on tracks generated from an astrophysically motivated power-law spectrum rather than mono-energetic or flat distributions. This results in a more realistic reconstruction setting, characterized by a larger fraction of low-energy, less informative tracks, and therefore a more challenging prediction task.}
\rev{It is also worth noting that, in the studies discussed above, the residual modulation obtained with the MoM on unpolarized datasets lies in a range ${<1\%}$, comparable to that achieved by the CNNs. In our setting, the MoM yields a significantly lower residual modulation, suggesting that its performance depends sensitively on the characteristics of the dataset, with our power-law setting potentially contributing to this behavior.} 
Within this context, our GNN \rev{exhibits} modulation levels comparable to \rev{those reported for} other ML-based methods. In contrast to the CNN architectures of~\cite{peir21,spie}, our model does not exhibit the pronounced harmonic artifacts at $n = 6$ and $n = 12$ observed in their Fourier spectra of reconstructed angles \cite{spie}. The achieved modulation is lower than that reported by~\cite{li25}, although their limited test sample prevents a statistically robust comparison. Finally, we \rev{obtain results comparable} to~\cite{Jiao2025}, despite using substantially less training data and without any post-hoc correction.

\begin{figure*}[h]
\centering
\includegraphics[scale=0.25]{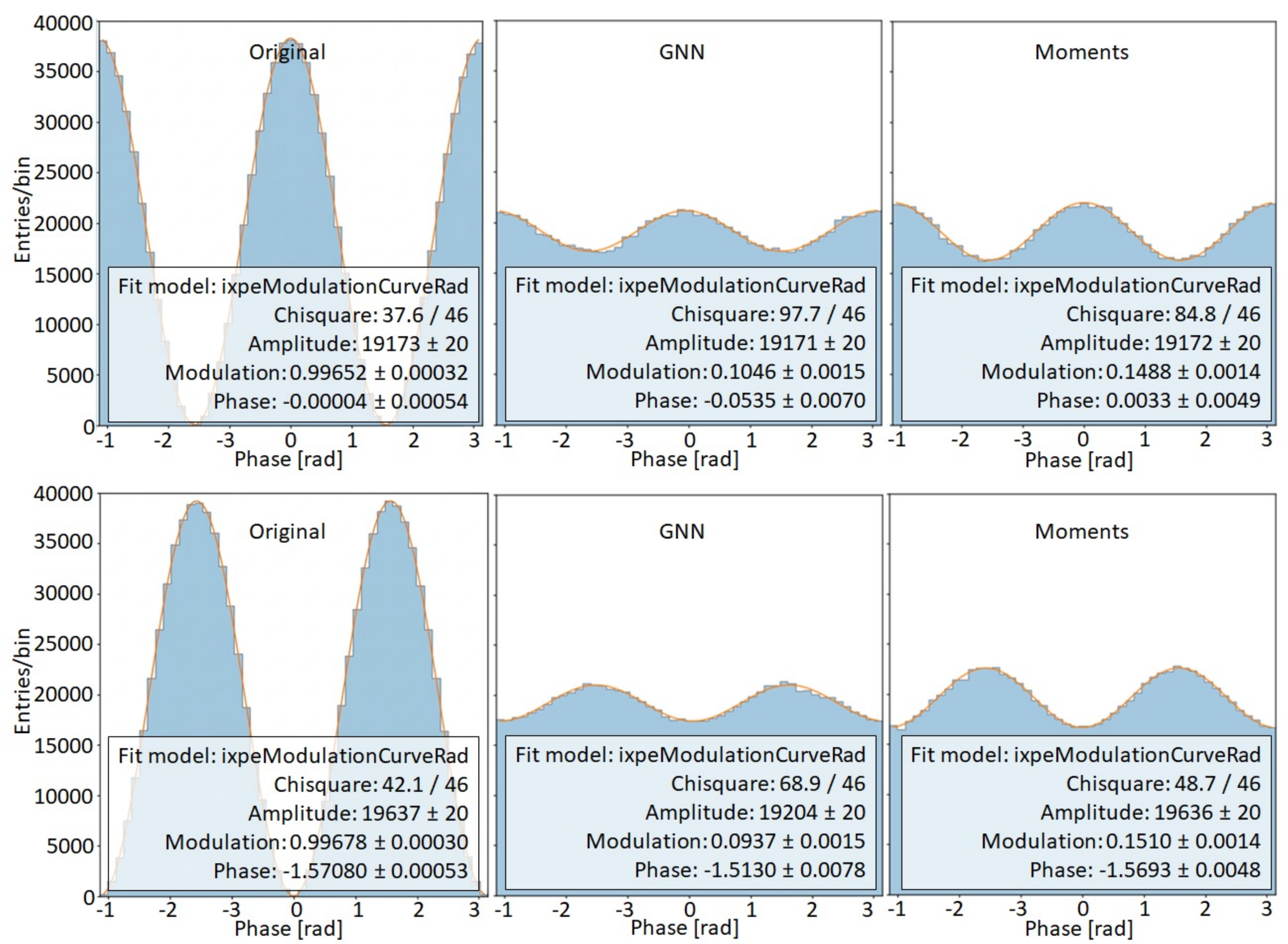}
\caption{Emission angle distributions \rev{original and reconstructed} for polarized data over one million tracks. The first row shows a signal following a cosine shape; the second row shows the same signal shifted by $90^\circ$. Original emission angles are shown on the left, GNN reconstructions in the center, and moment method reconstructions on the right.}
\label{fig: pol}
\end{figure*}

For the evaluation of polarized datasets, we present results of two configurations to assess whether the trained neural network is capable of reconstructing previously unseen polarization signals. 
To verify this, we test the network on two distinct polarized datasets: one in which the polarization follows a cosine modulation and the second where the signal is phase-shifted by $90^\circ$.

\rev{The results are shown in Figure~\ref{fig: pol}. The MoM reconstructs both the phase and the modulation of the signal with high accuracy, recovering the expected peak and trough positions with a phase shift of approximately $0.2^\circ$. 
The GNN, while yielding a lower modulation amplitude, is still able to reconstruct the overall structure of the polarization pattern in both the $\cos^2$ and $90^\circ$-shifted $\cos^2$ configurations, correctly identifying the locations of peaks and troughs, with a phase shift of about $3^\circ$. 
Notably, the GNN is trained exclusively on unpolarized data and has not been exposed to polarized signals during training, making this result a test of its ability to generalize beyond the training distribution.}

\rev{It is also worth noting that even the MoM does not fully recover the ideal modulation amplitude. This systematic underestimation arises from a loss of directional information, due to the diffusion of primary electrons during their drift toward the readout plane, which effectively blurs the track. This indicates that accurate modulation reconstruction is intrinsically challenging, leaving room for further improvements beyond current approaches.
In this context, the objective of reconstruction methods is not to reproduce the ideal angular distribution exactly, but to reliably detect the presence of a polarization signal and recover its key characteristics, namely the modulation amplitude and phase. 
The GNN is able to capture the underlying polarization pattern, despite the attenuation of the modulation, while the MoM achieves a more accurate reconstruction.}

\rev{Regarding CNN-based approaches, previous works report competitive or improved modulation performance compared to the MoM under specific conditions.
However, these comparisons should be interpreted with caution because our power-law setting includes a larger fraction of low-energy, less informative tracks, resulting in a more demanding reconstruction scenario. Furthermore, in some cases \cite{peir21,spie}, the CNNs introduce bias in the response to unpolarized sources.}

\rev{The observed behavior of the GNN can be attributed to factors related to the training strategy and the nature of the reconstruction task.
First, the network is trained exclusively on unpolarized data, which is effective in suppressing spurious modulation, but may also reduce the model's sensitivity, leading to an attenuation of the reconstructed modulation on polarized data. Second, the loss function is defined at the level of individual events, comparing predicted and true angles through their sine and cosine components. This formulation optimizes local reconstruction accuracy, but does not directly constrain global observables such as the modulation amplitude, which emerge only after aggregating a large number of events. As a result, per-event deviations induce a smoothing of the reconstructed angular distribution and a systematic underestimation of the modulation amplitude.
Taken together, these factors highlight a non-trivial relationship between per-event reconstruction accuracy and the recovery of global observables, motivating the analysis of prediction errors at the event level presented in the following subsection.}


\subsection{ Prediction Error}
In this subsection, we report the mean squared error (MSE) computed on the
angles reconstructed by the GNN model and the corresponding true emission angles. For comparison, we also compute the MSE for the classical MoM under the same conditions. 
\rev{Prior studies primarily focus on polarimetric sensitivity such as maximizing the modulation factor and minimizing  the minimum detectable polarization, and only rarely report per-track angular reconstruction errors.}

\rev{However, the considered task involves two distinct levels: a local level, concerned with the per-track angular reconstruction error, and a global level, related to the reconstruction of the overall angular distribution, from which the presence and phase of polarization are inferred. 
Therefore, we include the MSE as an additional metric to quantify the fidelity of per-event reconstruction and to analyze its relationship with global observables such as the modulation. This allows us to assess how improvements at the level of individual predictions relate to the reconstruction of the angular distribution.}

\begin{table*}[h]
\centering
\caption{Mean squared error (MSE) between reconstructed and true emission angles for the GNN and moment method, binned by source energy and for each polarization configuration.}
\begin{tabular}{c|cc|cc|cc}													
\toprule													
\textbf{Energy Bin (keV)} 	&	 \multicolumn{2}{c|}{\textbf{Flat}} 	&	 \multicolumn{2}{c|}{\textbf{Polarized}} 	&	 \multicolumn{2}{c}{\textbf{Polarized + 90°}} \\							\\
 	&	 GNN 	&	 MoM 	&	 GNN 	&	 MoM 	&	 GNN 	&	 MoM 	\\
\midrule													
0–2     	&	0.8956	&	0.9064	&	0.8904	&	0.9079	&	0.8978	&	0.9071	\\
2–3     	&	0.8956	&	0.9090	&	0.8944	&	0.9118	&	0.8990	&	0.9088	\\
3–4     	&	0.8962	&	0.9084	&	0.8975	&	0.9098	&	0.8999	&	0.9046	\\
4–5     	&	0.8960	&	0.9057	&	0.8943	&	0.9092	&	0.8945	&	0.9061	\\
5–6     	&	0.8953	&	0.9063	&	0.8840	&	0.8877	&	0.8943	&	0.9077	\\
6–7     	&	0.9080	&	0.9043	&	0.8921	&	0.9105	&	0.9128	&	0.8908	\\
7–8     	&	0.8855	&	0.9179	&	0.8790	&	0.9142	&	0.8968	&	0.9002	\\
8–inf   	&	0.8855	&	0.9021	&	0.8750	&	0.8995	&	0.9069	&	0.9207	\\
\midrule													
\textbf{All energies} 	&	0.8952	&	0.9070	&	0.8919	&	0.9087	&	0.8980	&	0.9070	\\
\bottomrule													
\end{tabular}

\label{tab: mse}
\end{table*}

The MSE values are reported in Table~\ref{tab: mse}. In this table, the values of the error are  computed on the sine/cosine components of the angles, and the results are divided into energy intervals (in keV) to analyze the accuracy of the two methods across different energy ranges. We observe that both methods yield similar levels of error, with the GNN ensemble  achieving slightly lower MSE values in most energy bins. When considering the overall error across the entire energy spectrum, the GNN exhibits a modest improvement in per-event reconstruction accuracy across all tested configurations.

\begin{table}[htbp]

\centering
\caption{Comparison of per-event squared errors between the GNN and MoM methods across different datasets. Negative values indicate lower error for the GNN. The confidence level for the intervals is 95\% as estimated by the \textit{t-test}.}

\begin{tabular}{lccc}
\hline
Dataset & Mean diff (NN $-$ MoM) & 95\% CI & p-value \\
\hline
Flat & $-0.0118$ & $[-0.0134,\,-0.0102]$ & $6.9 \times 10^{-47}$ \\
Polarized & $-0.0164$ & $[-0.0180,\,-0.0148]$ & $5.9 \times 10^{-91}$ \\
Polarized $+\,90^\circ$ & $-0.0091$ & $[-0.0107,\,-0.0074]$ & $1.0 \times 10^{-27}$ \\
\hline
\end{tabular}
\label{tab:statistical_significance}
\end{table}

\rev{To quantify the statistical significance of these differences, we perform a paired analysis on the per-event squared errors between the GNN and MoM methods. The results, reported in Table~\ref{tab:statistical_significance}, show that the GNN consistently achieves lower errors across all configurations. The mean differences (NN $-$ MoM) are negative in all cases, and the corresponding 95\% confidence intervals do not include zero. Paired t-tests confirm that these differences are statistically significant, with p-values well below conventional thresholds \revv{such as $0.05$ and $0.01$}. These results indicate that the observed differences in MSE are systematic rather than due to statistical fluctuations.}

\rev{The results on the MSE highlight a key finding of this study: improved accuracy at the level of individual tracks does not necessarily translate into a better recovery of the overall modulation pattern. This reveals a disconnect between per-track angular reconstruction error and the reconstruction of the global angular distribution. 
This behavior emphasizes the importance of jointly analyzing local reconstruction accuracy and global performance metrics. However, these results leave open the question of whether, at sufficiently low reconstruction errors, per-track accuracy and global performance become more strongly correlated.}



\section{Conclusions}
In this paper, we \rev{present a study of a Graph Neural Network approach} for the challenging task of reconstructing the polarization waveform of light beams from photoelectron tracks recorded by a gas pixel detector. 
We adopt a GNN approach because the underlying detector data are defined on a hexagonal pixel mesh and exhibit sparse activation.
\rev{The GNN is compared with the Method of Moments, the standard approach for this problem, to analyze the advantages and limitations of neural networks in this setting.}

{To minimize modeling bias, we train the network exclusively on simulated tracks from an unpolarized source. We employ rotational data augmentation and ensemble selection to improve angular generalization and mitigate structural artifacts. Thanks to the graph-based representation, these transformations can be applied without introducing distortions. At inference time, predictions are aggregated across multiple transformations, and models with the lowest residual modulation on unpolarized data are selected to further suppress spurious signals.}


\rev{In contrast to previous works, the GNN is trained and tested on tracks generated from an astrophysically motivated power-law photon spectrum, rather than on monoenergetic datasets, resulting in a more realistic and challenging reconstruction setting. Our results show that the MoM achieves a better modulation than the GNN for both unpolarized and polarized sources.
}
\rev{The analysis of the mean squared error (MSE) at the per-track level, computed on the sine and cosine components of the reconstructed angles, reveals a counterintuitive result: although the MoM exhibits better modulation, it consistently achieves a worse per-track reconstruction error than the GNN. This finding shows that improvements in per-track angular reconstruction do not necessarily translate into a better reconstruction of the overall modulation pattern.
}
\rev{
Notably, even the MoM does not reach the ideal modulation amplitude in polarized datasets, highlighting the intrinsic difficulty of the reconstruction task.}

\rev{A key direction for future work is the validation of the GNN on real detector data, to assess its robustness under realistic conditions.}
We plan to apply the trained GNN ensemble to IXPE ground calibration and in-orbit observations to study domain shifts between simulated and measured tracks and to evaluate its robustness against detector noise and non-uniformities. This validation will help determine whether the low residual modulation observed in simulations translates into reliable polarization measurements, and clarify how the GNN framework can complement the standard Method of Moments in IXPE analyses.

\section*{Acknowledgments}
The authors thank Giancarlo Baglioni for managing the Ipazia server, which was used for the computational work presented in this paper. We thank Alessandra Celletti and Ugo Locatelli for facilitating our collaboration. The authors acknowledge the support of the Italian Space Agency (ASI) for the organization in the early stages of this project and for funding. We also acknowledge the encouraging support of Barbara Negri (ASI). We also would like to thank the anonymous reviewers for their insightful comments and constructive suggestions, which have significantly helped in improving the clarity and the focus of the manuscript.

\appendix
\rev{
\section{Definitions of Physical Quantities and Observables}\label{app:phys}
In this appendix, we provide brief definitions of the application-specific terms and physical quantities used throughout the paper, including those appearing in the figures.}

\begin{itemize}

\item\rev{\textbf{ Power-law spectrum:}  The power-law spectrum represents the characteristic spectrum of many celestial sources. 
In particular, the Crab Nebula---the archetypal X-ray source---exhibits a spectrum well 
described by a power law with photon index 
\[
N_{\text{phot}} \propto E^{-2}.
\]
This emission arises from a power-law distribution of relativistic electrons, such as those 
accelerated at the shock front produced by the neutron-star wind interacting with the surrounding 
interstellar medium.
Similarly, Active Galactic Nuclei (AGN) commonly display comparable spectra, typically generated 
through inverse-Compton scattering of UV photons emitted by the accretion disk in a hot corona 
surrounding the central supermassive black hole.}

\item\rev{\textbf{ Polarization:} A photon is characterized by its polarization, which can be linear or circular. In this work, we consider only linear polarization, as it is the component measurable by our detector.}
\item\rev{\textbf{ Photoelectron (PE):} When an X-ray photon interacts with a gas atom, it ejects a photoelectron.}

\item\rev{\textbf{Photoelectron Emission:}  The emission direction of a PE is not isotropic, instead it is preferentially aligned with the electric field vector of the incident photon (its linear polarization direction). In the case of $s$-shell photoelectrons, emission is forbidden in directions perpendicular to the polarization vector; thus, the emission pattern is fully modulated. If an incident beam has a polarization degree $f$, the distribution of emission angles is modulated with the same fraction $f$.}
\item\rev{ \textbf{Photoelectron Emission Angle:} the angle between the photoelectron emission direction and the photon polarization vector. }

\item\rev{\textbf{Photoelectron Track:} As the PE propagates through the gas medium, it produces a trail of electron–ion pairs along its path. While this track encodes the original emission direction, occasional scattering with atomic nuclei causes abrupt deviations, leading to a loss of directional information. The resulting sequence of ionization events forms the \textit{photoelectron track}, which—after being blurred by diffusion during the drift toward the sensor and discretized into detector pixels—serves as the input to our reconstruction models. The emission direction of the photoelectron (PE), as inferred from the tracks recorded by the detector, may differ from the true initial direction, leading to a reduction of the \textit{modulation} relative to that obtained from the original emission directions. This reduction is expected to be more pronounced at lower energies, where tracks are shorter, diffusion effects are stronger, and elastic scattering with nuclei is more efficient.}

\item\rev{\textbf{Drift and Diffusion:} The track is formed almost instantaneously and is driven by an electric field toward the sensitive readout plane. During this drift (typically over a distance of $\sim$0.5~cm), the primary electrons undergo diffusion due to thermal agitation and collisions with gas molecules, which spatially blurs the track.}

\item\rev{\textbf{Signal Amplification (GEM):} Before reaching the readout, the track passes through a Gas Electron Multiplier (GEM). Within its microscopic holes, an intense electric field triggers avalanche multiplication, producing approximately 150–200 secondary electrons per primary electron. This process increases the signal-to-noise ratio enough for the pixelated readout plane to detect the track with high fidelity without significantly altering its morphology.}

\item\rev{\textbf{Chisquare:} The chi-squared statistic is defined as $\chi^2 = \sum_{i=1}^{n} \frac{(O_i - E_i)^2}{E_i}$, where $O_i$ and $E_i$ denote the observed and expected values, respectively. The second value corresponds to the number of degrees of freedom. A ratio close to unity indicates that the fitting function provides an adequate description of the data.}

\end{itemize}

\printbibliography

\end{document}